\documentclass[reprint,showpacs,twocolumn,preprintnumbers,amsmath,amssymb,aip,graphicx,apl]{revtex4-1}

\usepackage{graphicx}
\usepackage{bm}
\usepackage[mathlines]{lineno}
\usepackage{ulem}
\usepackage{color}
\usepackage{comment}

\begin{document}

\title[Short Title]{Enhancement of Coherent Phonon Amplitude in Phase-change Materials by Near-infrared Laser Irradiation
}

\author{Takara Suzuki}
\email{s1720369@u.tsukuba.ac.jp}
\affiliation{Division of Applied Physics, Faculty of Pure and Applied Sciences, University of Tsukuba, 1-1-1 Tennodai, Tsukuba 305-8573, Japan.}
\author{Yuta Saito}
\author{Paul Fons}
\author{Alexander V. Kolobov}
\author{Junji Tominaga}
\affiliation{Nanoelectronics Research Institute, National Institute of Advanced Industrial Science and Technology,
Tsukuba Central 5, 1-1-1 Higashi, Tsukuba 305-8565, Japan}
\author{Muneaki Hase}
\email{mhase@bk.tsukuba.ac.jp}
\affiliation{Division of Applied Physics, Faculty of Pure and Applied Sciences, University of Tsukuba, 1-1-1 Tennodai, Tsukuba 305-8573, Japan.}

\date{\today}

\begin{abstract}
We have examined the effect of pump-probe photon energy on the amplitude of coherent optical phonons in a prototypical phase change material using a femtosecond time-resolved transmission technique. The photon energy was varied between 0.8 and 1.0 eV (corresponding to the wavelength of 1550 and 1200 nm), a range over which there is significant optical contrast between the crystalline and amorphous phases of Ge$_{2}$Sb$_{2}$Te$_{5}$ (GST225). It was found that in crystalline GST225 the coherent phonon amplitude monotonically increases as the photon energy increases, indicating that the phonon amplitude is enhanced by interband optical absorption, which is associated with the imaginary part of the dielectric function. In amorphous GST225, in contrast, the coherent phonon amplitude does not depend on the photon energy, remaining almost constant over the tuning range. A possible contribution from the polarizability associated with resonant bonding nature of GST225 is discussed. 
\end{abstract}

\maketitle
Modern optical data storage technology was developed based upon phase change materials (PCMs), which exhibit a large optical contrast between the crystalline and amorphous states.\cite{Yamada91,Wuttig07} Ge$_{2}$Sb$_{2}$Te$_{5}$ (GST225) is one of the best-performing chalcogenide alloys and has been utilized as the recording material in Digital Versatile Disc Random Access Memory (DVD-RAM) since it exhibits a large optical contrast, typically more than $\approx$ 20 \%.\cite{Wuttig07,Siegel04} The optical properties of DVD-RAM have been extensively investigated experimentally and theoretically, where optical absorption occurs from the visible to the near-infrared wavelength region due to the narrow band-gap energy range of 0.5 - 0.75 eV.\cite{Lee04} For crystalline GST225 the real part of the dielectric function $\varepsilon_{1}$ has a maximum at $\approx$ 0.7 eV, while the imaginary part $\varepsilon_{2}$ has a maximum at 1.2 eV.\cite{Lee04,Orava08} For the case of amorphous GST225, on the other hand, $\varepsilon_{1}$ has a maximum at $\approx$ 1.2 eV, while $\varepsilon_{2}$ has a maximum at 1.9 eV.\cite{Lee04,Orava08}  For GST225, in the crystalline phase the real part $\varepsilon_{1}$ is usually associated with resonant-bonding and the imaginary part $\varepsilon_{2}$ is associated with interband photo-absorption.\cite{Shportko08} When the laser photon energy is tuned through these critical energies, the resonant excitation of coherent phonons via either {\it resonant} impulsive stimulated Raman scattering (ISRS)\cite{Stevens02} or displacive excitation of coherent phonon (DECP)\cite{Zeiger92} is expected to occur. The photon energy dependence of the coherent phonon amplitude excited via either {\it resonant} ISRS or DECP can be described by the two-tensor model, which includes $\varepsilon_{1}$ and $\varepsilon_{2}$.\cite{Stevens02}

It is unusual to investigate the dependence of the amplitude of phonon spectra with laser photon energy; typically the frequency shift has been studied as a function of pump fluence.\cite{Makino11,Miller16,Rueda11}
In this paper, we have investigated the pump-probe wavelength dependence of coherent phonon excitation in GST225 using a transmission pump-probe technique [Fig. 1(a)]. 
Between 1200 nm and 1550 nm (1.0 and 0.8 eV), we found that the shorter the wavelength, the larger the coherent phonon oscillation amplitude became. The Fourier transformed spectra exhibited a dominant peak at 3.2 THz, and the increase of this peak can be well explained by the dominant contribution from interband photo-absorption associated with the imaginary part of the dielectric function.\cite{Lee04,Orava08}

The samples used in the present study were thin films of Ge$_{2}$Sb$_{2}$Te$_{5}$ (GST225) fabricated using helicon-wave RF magnetron sputtering onto Si(111) substrates, whose thickness was 100 $\mu$m. The GST225 films of 80-nm-thickness were grown by co-sputtering of GeTe and Sb$_{2}$Te$_{3}$ alloy targets. By annealing the as deposited amorphous GST225 films at $280\ {}^\circ\mathrm{C}$ for 1 hour, polycrystalline (face-center-cubic: fcc) GST225 films were also obtained.\cite{Forst00,Friedrich00} A ZnS-SiO$_{2}$ (20 nm) layer was deposited on the GST225 film to prevent oxidation. X-ray diffraction from GST films, which were prepared under the same conditions, confirmed the structural transition of the film from the amorphous to the crystalline phase. An additional piece of the Si(111) substrate was also used as a reference sample, to check the time resolution of the pump-probe system in the present study based on two-photon absorption (TPA).\cite{Tsang02} \\
\begin{figure}
\includegraphics[width=86mm]{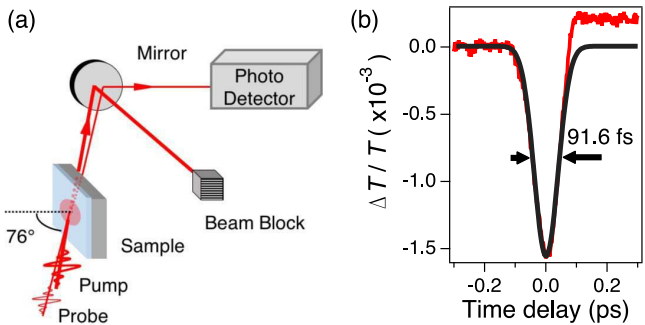}
\caption {(a) Experimental configuration around the sample (GST225). The angle of incidence has been adjusted to the Brewster angle ($\approx 76^\circ$) to efficiently excite the sample. (b) Time-resolved transmission change in the Si(111) substrate (red) and its Gaussian fit (black). Transient reduction in the transmission is caused by two photon absorption (TPA), whose FWHM is estimated to $\approx$ 91.6 fs by the Gaussian fit. }
\label{Fig. 1}
\end{figure}
In the experiment, we utilized a transmission pump-probe technique to obtain a larger oscillation intensity than that obtainable using reflective techniques.\cite{Miller16} The pump-probe set up used is illustrated in Fig. 1(a). Femtosecond laser pulses from a mode locked Ti:sapphire laser (pulse width 20 fs, central wavelength 800 nm, repetition rate 80 MHz, average power 525 mW) were amplified by a regenerative amplifier system RegA9040, to obtain 40 fs laser pulses at an average power of 590 mW at 100 kHz. After that the output laser pulses were delivered to an optical parametric amplifier (OPA9850) to generate infrared laser pulses, whose wavelength could be tuned from 1200 to 1550 nm. The angle of incidence used with respect to the sample surface was decided from its Brewster angle ($76^\circ$) to allow the $p$-polarized pump and probe pulses propagate through the sample efficiently. Before each experiment, the BBO crystal in the OPA9850 was rotated to generate a given wavelength output while being monitored by a spectrometer. The pump and probe wavelength were identical in the present experiment. 
The photo-induced transmission change ($\Delta${\it T}/{\it T}) was thus recorded as a function of the time delay between the pump and probe pulses. The delay was scanned over 10 ps and averaged for 5,000 scans, using an oscillating retroreflector with a 10 Hz scan frequency.\cite{Hase12} The pump power was kept constant at 30 mW over the wavelength range used, which corresponds to a fluence of $\approx$ 1.4 mJ/cm$^{2}$, a value calculated using the focused ellipsoidal beam diameter of $\approx 70\times$289 $\mu$m (the spot area of 70$\times$70 $\mu$m at the normal incident angle was measured using a knife-edge) due to the large incident angle of $\approx 76^\circ$. The optical penetration depth of $\approx$ 100 nm at 1550 nm, which was calculated from the absorption coefficient,\cite{Lee04} suggests negligibly small inhomogeneous excitation effects along the sample depth. 
We have carefully examined multi-reflections originating from the GST/Si interface inside the sample, and found that Snell$'$s law does not allow effective multi-reflection. In the crystalline sample, the refractive index $n$ is $\approx$ 6.8 and is nearly constant over the entire wavelength range used.\cite{Lee04} Based upon this, the penetration depth is estimated to be $\approx$ 100 nm, which is much shorter than the distance that the first order multi-reflection beam must propagate through the sample, $\approx$ 242 nm. In addition, the first reflection at the GST/Si interface is calculated to be only $\approx$ 10 \%, making the reabsorption of the pump beam inside the sample negligible. 

Figure 1(b) shows the transient transmission $\Delta${\it T}/{\it T} obtained from a Si(111) substrate acquired using 1550 nm photons. Since the band-gap energy of Si is 1.11 eV ($\approx$ 1117 nm) at room temperature,\cite{Yu05} the ultrafast response observed for time delay zero is due to nonlinear optical effects, which are dominated by TPA.\cite{Tsang02} In the current case, we can extract the pulse length of the 1550 nm photon as $\approx$ 65 fs by using the full width at half maximum (FWHM = 91.6 fs) of the TPA signal based on a fit with a Gaussian function (91.6/1.414 $\approx$ 65 fs).\cite{Tsang02} We used CaF$_{2}$ substrates and lenses, which have nearly zero group delay dispersion between 1200 and 1550 nm, for our pump-probe optics. Therefore, the temporal resolution of the experiment did not change with wavelength.
\begin{figure}
\includegraphics[width=85mm]{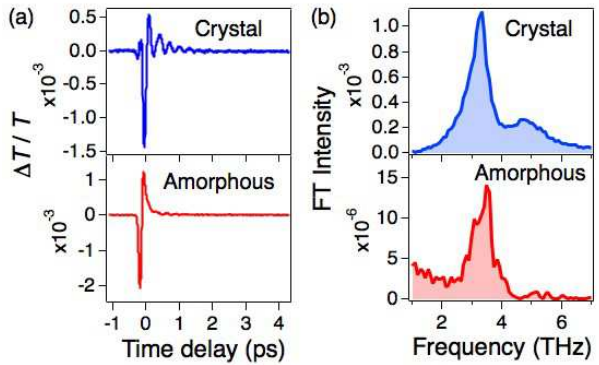}
\caption {Pump-probe experimental results, obtained for crystalline and amorphous GST225 thin films. (a) Time-resolved transmission change as a function of the time delay for two different phases obtained at 1400 nm. (b) Their FT spectra, which are contrasted by the two different phases at the local structure level.}
\label{Fig. 2}
\end{figure}
The transient transmission $\Delta${\it T}/{\it T} at 1400 nm obtained from amorphous and crystalline GST225 thin films is displayed in Fig.  2(a). The initial transient signal is dominated by the electronic response from GST225, since the optical absorption is much stronger in GST225 than that in the Si substrate at 1400 nm, which is followed by coherent oscillations. Coherent phonons can be clearly observed in the crystalline phase, but in the amorphous phase oscillations are not obvious. Fourier transformed (FT) spectra were obtained from the time-domain data as shown in Fig. 2(b). Since the phonon oscillation signal was very small in the amorphous phase, we have fit the time-domain data to an exponentially decaying function and subtracted the electronic response.\cite{Hase05} 
After this procedure, only the residual oscillatory signal was Fourier transformed, and the result is displayed in the figure. The FT spectra exhibit different spectral features in the two phases similar to those observed at much shorter wavelengths.\cite{Forst00,Miller16,Makino11,Rueda11} The main peak observed in the crystalline phase was 3.2 THz, which is ascribed to the A$_{1}$ mode,\cite{Forst00,Andrikopoulos07} while the main peak in the amorphous phase was located at 3.5 THz; A similar 0.3 THz peak shift was observed in GST225 under strong photoexcitation using a 1.55 eV (800 nm) photon pump.\cite{Rueda11,Hase15} 
Note that the initial phase of the excited optical phonon is {\it cosine-like}, indicating the generation mechanism is either DECP or resonant ISRS.\cite{Ishioka08,Misochko16} 

The transient transmission $\Delta${\it T}/{\it T} observed in crystalline GST225 for various wavelengths is displayed in Fig. 3(a), together with the corresponding FT spectra in Fig. 3(b). The transient $\Delta${\it T}/{\it T} signal became smaller as the wavelength was tuned from 1200 up to 1550 nm. The data for amorphous GST225 are not shown since its coherent phonon spectra does not significantly depend on the wavelength as discussed in Fig. 4. Figure 3(b) demonstrates the strong wavelength dependence of the coherent phonon spectra in the crystalline phase of GST225. The intensity of the main peak at 3.2 THz is clearly enhanced as the wavelength of the laser becomes shorter. On the other hand, upon decreasing the wavelength, a red-shift in the peak and an associated broadening in the spectral width (FWHM) are observed for the 3.2 THz A$_{1}$ mode, which can be explained by strong electron-phonon coupling and/or photo-induced temperature effects; both effects are well known to occur when the phonon amplitude increases.\cite{Makino11,Rueda11} The wavelength dependence of the phonon amplitude in the infrared region, however, has not yet been explored. 
\begin{figure}
\includegraphics[width=88mm]{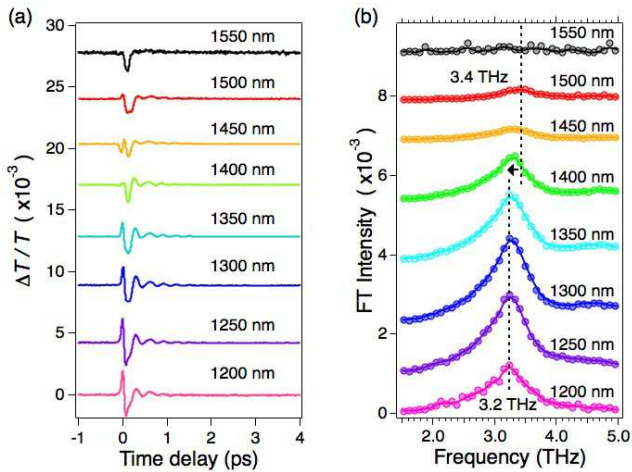}
\caption {Time-resolved transmission changes for crystalline GST225 film, pump and probed at various center wavelengths, are shown in (a) together with their FT spectra in (b). A red-shift in the peak frequency for the A$_{1}$ mode, from 3.4 THz down to 3.2 THz, is observed when the pump wavelength was varied from 1500 nm to 1200 nm. 
}
\label{Fig. 3}
\end{figure}
\nopagebreak

\begin{figure}
\includegraphics[width=87mm]{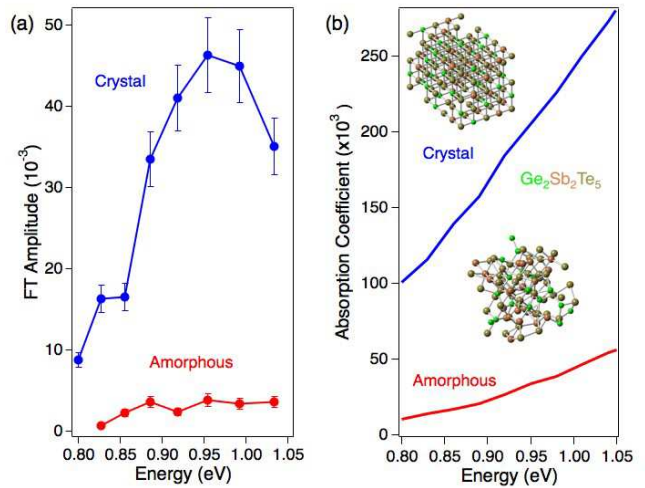}
\caption {(a) The peak FT amplitude of the A$_{1}$ mode plotted as a function of the photon energy of the pump and probe pulses converted from the center wavelength. (b) The absorption coefficient of Ge$_{2}$Sb$_{2}$Te$_{5}$ for the amorphous and crystalline phases, as a function of the photon energy. The data (the absorption coefficient) are taken from Ref. [4] and plotted. The inset shows the structure of the two phases drawn by CrystalMaker, in which Ge atoms are shown in green, Te atoms - in yellow, and Sb atoms are shown in orange.} 
\label{Fig. 4}
\end{figure}
The FT amplitude of the A$_{1}$ mode in the crystalline phase GST225 at 3.2 THz is plotted in Fig. 4(a) as a function of photon energy, together with the 
absorption coefficient in Fig. 4(b). The FT amplitude of the A$_{1}$ mode in crystalline GST225 increases nearly monotonically
(but decreases at $>$ 1.0 eV photon energy) with increasing photon energy, while that of the amorphous phase does not significantly depend on photon energy. The similarity of the dependence of the coherent phonon amplitude and the 
absorption coefficient on photon energy implies that the generation mechanism of the coherent phonon is governed by the so-called two tensor model.\cite{Stevens02,Garrett96,Hase05,Misochko16}

As shown in Fig. 4, both the FT amplitude and the 
absorption coefficient are enhanced when the photon energy becomes larger. It was also found that the contrast of the FT amplitude for the two phases in Fig. 4(a) coincides with that of the 
absorption coefficient shown in Fig. 4(b), implying that interband transitions, which increase toward $\approx$ 1.2 eV, play a major role in the enhancement of the FT amplitude of the coherent A$_{1}$ mode. 
The fluence dependence of the FT amplitude performed at 1300 nm (0.95 eV) implies that the FT amplitude increases in a  similar manner with the pump fluence. 
Additionally, the photon energy dependence of the FT amplitude should follow from the unique structure of crystalline GST225 [see the inset of Fig. 4(b)], especially, the electronic band structure resulting from resonant bonding.\cite{Shportko08} The band gap energy of the crystalline phase is $\approx$ 0.5 eV, while that of the amorphous is 0.7 eV,\cite{Lee04} leading to the difference in optical contrast between the two phases, e.g., the effect of the change in band gap and the absorption coefficient. 

Resonant bonding is an important characteristic of crystalline GST225, in which octahedrally coordinated bonding occurs even in the presence of only three $p$-electrons/atom leading to the strong polarizability of the valence electron charge.\cite{Shportko08} The interatomic distance arising from resonant bonding is longer and interaction is weaker than the corresponding covalent bond in the amorphous phase, which may result in the larger phonon oscillation amplitude observed. In the present experiment, we did not completely induce the collapse of the resonant bonds with a pump fluence of 1.4 mJ/cm$^{2}$,\cite{Waldecker15}  and hence resonant bonding was preserved in the crystalline GST225 sample. A possible effect of resonant bonding on the photon energy dependence of the FT amplitude is discussed from the Raman tensor point of view below.

In resonant-type ISRS, the reflectance change $\Delta${\it R}/{\it R} or transmission change $\Delta${\it T}/{\it T} is proportional to the elements of the Raman tensor ($\partial\chi/\partial${\it Q}), which is given by the  two stimulated Raman tensors, $\chi^{R}$ and $\pi^{R}$. The first tensor $\chi^{R}$ ($\propto$ $d\varepsilon_{1}/d\omega + i d\varepsilon_{2}/d\omega$) is the standard Raman tensor and is equivalent to the Raman polarizability $\alpha$,\cite{Dekorsy00} while the second tensor is 
the electrostrictive force acting on the ions, expressed by $\pi^{R}$$\propto$ $d\varepsilon_{1}/d\omega + 2i \varepsilon_{2}/\Omega$ with $\Omega$ being the phonon frequency and $\omega$ the laser frequency.\cite{Stevens02} Thus, we have approximately $\Delta${\it T}/{\it T} $\approx$ 1/{\it T}($\partial${\it T}/$\partial\chi$)($\partial\chi$/$\partial${\it Q})$\Delta${\it Q} $\approx$ 1/{\it T}($\partial${\it T}/$\partial\chi$)$\alpha\times\pi^{R}$. 

In order to examine the effect of the probe photon energy,\cite{Novelli17} we performed an additional experiment, with a 800 nm pump and an OPA probe in a reflection geometry [Figs. 5(a) and (b)]. 
The pump power was fixed at 30 mW, corresponding to a fluence of $\approx$ 1.4 mJ/cm$^{2}$, which is the same pump fluence as in the single-color experiment. 
In the experiment, we observed $\sim$ 200 \% enhancement of the FT amplitude for the crystalline GST225 sample, depending on the probe photon energy, from 0.83 to 1.0 eV [see Fig. 5(c)]. 
Compared with the FT amplitude vs photon energy in Fig. 4(a), the probe photon energy dependence qualitatively explains the enhancement of the FT amplitude in Fig. 4(a), while the magnitude 
of the enhancement cannot be fully explained by the probe photon energy dependence. We think that the combination of the pump-probe actions, i.e., $\pi^{R}$: pump and $\alpha$: probe, would produce a $\sim$ 300 \% enhancement of the FT amplitude in Fig. 4(a). 

Regarding the pumping action, the Raman tensor ($\pi^{R}$) is dominated by the imaginary part of the dielectric constant $\varepsilon_{2}$ (corresponding to the absorption coefficient) rather than the real part $\varepsilon_{1}$ under the condition that $|d\varepsilon_{1}/d\omega| \ll \varepsilon_{2}/\Omega$,\cite{Stevens02} or equivalently the real component $\varepsilon_{1}$ does not change dramatically as a function of the laser frequency $\omega$. This is the case here. 
\begin{figure}
\includegraphics[width=87mm]{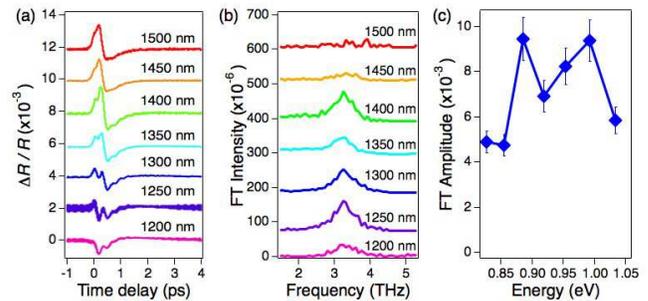}
\caption {(a) Time-resolved reflectivity changes for crystalline GST225 film with fixed pump (800 nm) and various probe wavelengths. (b) FT spectra of (a) are shown. We omit the FT spectrum at 1550 nm due to larger noise, making unable to compare with others. (c) The peak FT amplitude of the A$_{1}$ mode obtained from (b) plotted as a function of the probe photon energy. The amplitude becomes larger as the photon energy increases and at 1.0 eV ($\approx$ 1250 nm), it is about $\sim$ 200 \% larger than that at 0.83 eV ($\approx$ 1500 nm).} 
\label{Fig. 5}
\end{figure}
In addition to the experimental probe wavelength dependence of $\chi^{R}$ or $\alpha$ [Fig. 5(c)], we need to consider the effect of resonant bonding. 
The calculated electronic polarizability for GST225 is 108\% larger in the crystalline phase of GST225 than in the amorphous phase,\cite{Shportko08} taking into account the relative increase in the electronic polarizability $\varepsilon_{\infty}^{cryst}$/$\varepsilon_{\infty}^{amorph} - 1$, where $\varepsilon_{\infty}^{cryst}$ and $\varepsilon_{\infty}^{amorph}$ denote the optical dielectric constants in the crystalline and amorphous phases, respectively. This increased electronic polarizability in crystalline GST225 is believed to originate from resonant bonding.\cite{Shportko08} Taking into account the fact that the Raman polarizability $\alpha \approx(\varepsilon_{\infty} - 1)^{1/2}$,\cite{Lucovsky73} we expect $\alpha$ to be $\geq$10 \% larger for the case of the crystalline phase versus that for the case of the amorphous phase. 

As a result, the phononic contrast in Fig. 4(a) is governed mainly by the effects of the imaginary part $\varepsilon_{2}$ (corresponding to the absorption coefficient) on $\pi^{R}$ and the probe wavelength dependence on $\alpha$, together with a small correction due to the effect of resonant bonding on $\alpha$.

In summary, we have explored the effect of the pump-probe photon energy on the amplitude of coherent optical phonons in GST225 thin films using femtosecond time-resolved transmission measurements. The pump and probe photon energies were tuned between 0.8 and 1.0 eV (corresponding to the wavelength of 1550 and 1200 nm), to investigate the phononic contrast between the crystalline and amorphous phases. We found that in crystalline GST225, the coherent A$_{1}$ phonon amplitude was enhanced as the photon energy increased, while in the amorphous phase it did not significantly depend on the photon energy. The phononic contrast in the wavelength region monitored in the present study can be explained by mainly contributions from the two Raman tensors, associated with the imaginary part of the dielectric function, and a minor contribution from the polarizability associated with resonant bonding. Our results provide highly relevant fundamental knowledge for ultrafast optical data processing and for the next-generation of ultra-high-speed phase change random access memory (PCRAM) technologies in the optical communication wavelength region.


This work was supported by CREST (NO. JPMJCR14F1), JST, Japan.
We acknowledge Ms. R. Kondou for sample preparation.\\


\pagebreak


\begin{thebibliography}{99}

\bibitem{Yamada91}
N. Yamada, E. Ohno, K. Nishiuchi, and N. Akahira, J. Appl. Phys. \textbf{69}, 2849 (1991). 

\bibitem{Wuttig07}
M. Wuttig and N. Yamada, Nat. Mater. \textbf{6}, 824 (2007).

\bibitem{Siegel04}
J. Siegel, A. Schropp, J. Solis, and C. N. Afonso, Appl. Phys. Lett \textbf{84}, 2250 (2004).

\bibitem{Lee04}
B. S. Lee, J. R. Abelsona, S. G. Bishop, D. H. Kang, B. K. Chong, and K. B. Kim, J. Appl. Phys. \textbf{97}, 093509 (2005).

\bibitem{Orava08}
J. Orava, T. W\'{a}gner, J. \v{S}ik, J. P\v{r}ikry, M. Frumar, and  L. Bene\v{s}, J. Appl. Phys. \textbf{104}, 043523 (2008).

\bibitem{Shportko08}
K. Shportko, S. Kremers, M. Woda, D. Lencer, J. Robertson, and M. Wuttig, Nat. Mater. \textbf{7}, 653 (2008).

\bibitem{Stevens02}
T. E. Stevens, J. Kuhl, and R. Merlin, Phys. Rev. B \textbf{65}, 144304 (2002).

\bibitem{Zeiger92}
H. J. Zeiger, J. Vidal, T. K. Cheng, E. P. Ippen, G. Dresselhaus, and M. S. Dresselhaus, Phys. Rev. B {\bf45}, 768 (1992).

\bibitem{Makino11}
K. Makino, J. Tominaga, and M. Hase, Opt. Exp. \textbf{19}, 1260 (2011).

\bibitem{Miller16}
T. A. Miller, M. Rud\'{e}, V. Pruneri, and S. Wall, Phys. Rev. B \textbf{94}, 024301 (2016).

\bibitem{Rueda11}
J. Hernandez-Rueda, A. Savoia, W. Gawelda, J. Solis, B. Mansart, D. Boschetto, and J. Siegel, Appl. Phys. Lett. \textbf{98}, 251906 (2011).

\bibitem{Forst00}
M. F\"{o}rst, T. Dekorsy, C. Trappe, M. Laurenzis, H. Kurz, and B. Bechevet, Appl. Phys. Lett. \textbf{77}, 1964 (2000).

\bibitem{Friedrich00}
I. Friedrich, V. Weidenhof, W. Njoroge, P. Franz, and M. Wuttig, J. Appl. Phys. \textbf{87}, 4130 (2000).

\bibitem{Tsang02}
H. K. Tsang, C. S. Wong, T. K. Liang, I. E. Day, S. W. Roberts, A. Harpin, J. Drake, and M. Asghari, Appl. Phys. Lett \textbf{80}, 416 (2002).

\bibitem{Hase12}
M. Hase, M. Katsuragawa, A. M. Constantinescu, and H. Petek, Nat. Photon. \textbf{6}, 243 (2012).

\bibitem{Yu05}
P. Yu and M. Cardona in \textit{Fundamentals of Semiconductors} (Springer-Verlag: Berlin Heidelberg, 2005).

\bibitem{Hase05}
M. Hase, K. Ishioka, J. Demsar, J. Ushida, and M. Kitajima, Phys. Rev. B \textbf{71}, 184301 (2005).

\bibitem{Andrikopoulos07}
K. S. Andrikopoulos, S. N. Yannopoulos, A. V. Kolobov, P. Fons, and J. Tominaga, J. Phys. Chem. Sol. \textbf{68}, 1074 (2007).

\bibitem{Hase15}
M. Hase, P. Fons, K. Mitrofanov, A. V. Kolobov, and J. Tominaga, Nat. Commun. \textbf{6}, 8367 (2015).

\bibitem{Ishioka08}
K. Ishioka, M. Kitajima, and O. V. Misochko, J. Appl. Phys. \textbf{103}, 123505 (2008).

\bibitem{Misochko16}
O. V. Misochko and M. V. Lebedev, Phys. Rev. B \textbf{94}, 184307 (2016).

\bibitem{Garrett96}
G. A. Garrett, T. F. Albrecht, J. F. Whitaker, and R. Merlin, Phys. Rev. Lett. \textbf{77}, 3661 (1996).

\bibitem{Waldecker15}
L. Waldecker, T. A. Miller, M. Rud\'{e}, R. Bertoni, J. Osmond, V. Pruneri, R. Simpson, R. Ernstorfer, and S. Wall, Nat. Mater. \textbf{14}, 991 (2015).

\bibitem{Dekorsy00}
T. Dekorsy, G. C. Cho, and H. Kurz, in {\it Topics in Applied Physics: Light Scattering in Solids VIII}, edited by M. Cardona and G. G\"{u}ntherodt (Springer Verlag, 2000), Vol. 76, p. 169. 

\bibitem{Novelli17}
F. Novelli, G. Giovannetti, A. Avella, F. Cilento, L. Patthey, M. Radovic, M. Capone, F. Parmigiani, and D. Fausti, Phys. Rev. B \textbf{95}, 174524 (2017). 

\bibitem{Lucovsky73}
G. Lucovsky and R. M. White, Phys. Rev. B \textbf{8}, 660 (1973).

\end{thebibliography}
\end{document}